\documentclass[11pt]{elsart}
\pdfoutput=1 

\usepackage{epsfig}
\usepackage{rotating}
\usepackage{amssymb}
\usepackage{moreverb}
\usepackage{amsmath}
\usepackage{amsbsy}
\usepackage{enumerate}
\usepackage{bbm}
\usepackage{url}
\usepackage{cite}

\newcommand\sss{\mathchoice%
{\displaystyle}%
{\scriptstyle}%
{\scriptscriptstyle}%
{\scriptscriptstyle}%
}

\newcommand\as{\alpha_{\sss\rm S}}
\newcommand\mur{\mu_{\sss\rm R}}

\def\be{\begin{equation}}
\def\ee{\end{equation}}
\def\bea{\begin{eqnarray*}}
\def\eea{\end{eqnarray*}}

\def\ell{l}

\def\({\left(}
\def\){\right)}

\begin{document}

\bibliographystyle{JHEP}

\begin{frontmatter}


\title{Update of the Binoth Les Houches Accord for a standard interface
  between Monte Carlo tools and one-loop programs}

\thispagestyle{empty}

\author[a0]{S.~Alioli},
\author[a1]{S.~Badger},
\author[a2a]{J.~Bellm},
\author[a2]{B.~Biedermann},
\author[a3]{F.~Boudjema},
\author[a4]{G.~Cullen},
\author[a5]{A.~Denner},
\author[a6]{H.~van~Deurzen},
\author[a7]{S.~Dittmaier},
\author[a8]{R.~Frederix},
\author[a8,a10]{S.~Frixione},
\author[a9]{M.V.Garzelli},
\author[a2a]{S.~Gieseke},
\author[a14]{E.W.N.~Glover},
\author[a6]{N.~Greiner},
\author[a6]{G.~Heinrich},
\author[a10]{V.~Hirschi},
\author[a11]{S.~H\"oche},
\author[a12]{J.~Huston},
\author[a7]{H.~Ita},
\author[a13]{N.~Kauer},
\author[a14]{F.~Krauss},
\author[a6]{G.~Luisoni},
\author[a14]{D.~Ma\^{\i}tre},
\author[a15]{F.~Maltoni},
\author[a18]{P.~Nason},
\author[a17]{C.~Oleari},
\author[a16]{R.~Pittau},
\author[a20]{S.~Pl\"atzer},
\author[a21]{S.~Pozzorini},
\author[a22]{L.~Reina},
\author[a2a]{C.~Reuschle},
\author[a23]{T.~Robens},
\author[a6]{J.~Schlenk},
\author[a14]{M.~Sch{\"o}nherr},
\author[a7]{F.~Siegert},
\author[a6]{J.F.von Soden-Fraunhofen},
\author[a20]{F.~Tackmann},
\author[a24]{F.~Tramontano},
\author[a2]{P.~Uwer},
\author[a8]{G.~Salam},
\author[a8]{P.~Skands},
\author[a25]{S.~Weinzierl},
\author[a6]{J.~Winter},
\author[a1]{V.~Yundin},
\author[a26]{G.~Zanderighi},
\author[a15]{M.~Zaro}

\address[a0]{Lawrence Berkeley National Laboratory and University of California, Berkeley, CA 94720, USA}
\address[a1]{The Niels Bohr Institute,University of Copenhagen, DK-2100 Copenhagen, Denmark}
\address[a2a]{Karlsruhe Institute of Technology, 76131 Karlsruhe, Germany}
\address[a2]{Humboldt-Universit\"at zu Berlin, Institut f\"ur Physik, D-12489 Berlin, Germany}
\address[a3]{LAPTH, Universit\'e de Savoie and CNRS, F-74941 Annecy-le-Vieux, France}
\address[a4]{Deutsches Elektronen-Synchrotron DESY, Platanenallee 6, 15738 Zeuthen, Germany}
\address[a5]{Universit\"at W\"urzburg, Institut f\"ur Theoretische Physik und Astrophysik, D-97074 W\"urzburg, Germany}
\address[a6]{Max Planck Institute for Physics,  80805 Munich, Germany}
\address[a7]{Albert-Ludwigs-Universit\"at Freiburg, Physikalisches  Institut, D-79104 Freiburg, Germany}
\address[a8]{CERN/PH, CH--1211 Geneva 23, Switzerland}
\address[a9]{University of Nova Gorica, SI 5000 Nova Gorica, Slovenia}
\address[a10]{ITTP, EPFL,  CH-1015 Lausanne, Switzerland}
\address[a11]{SLAC, Stanford University, Stanford, CA 94309, USA}
\address[a12]{Michigan State University, East Lansing, MI 48840, USA}
\address[a13]{Department of Physics, Royal Holloway, University of London, Egham TW20 0EX, UK}
\address[a14]{Institute for Particle Physics Phenomenology, University of Durham, Durham, DH1 3LE, UK}
\address[a15]{CP3, Universit\'{e} Catholique de Louvain,1348 Louvain-la-Neuve, Belgium}
\address[a16]{Departamento de F´isica Te´orica y del Cosmos CAFPE, Universidad de Granada, E-18071 Granada, Spain}
\address[a17]{Universit\`a di Milano-Bicocca and INFN, Sezione di Milano-Bicocca, 20126 Milano, Italy}
\address[a18]{INFN, Sezione di Milano-Bicocca, 20126 Milano, Italy}
\address[a20]{DESY Hamburg, Germany}
\address[a21]{University of Zurich, Institute for Theoretical Physics,  CH-8057 Zurich, Switzerland}
\address[a22]{Florida State University, Tallahassee, FL 32306-4350, USA}
\address[a23]{Technical University Dresden, 01062 Dresden, Germany}
\address[a24]{Dipartimento di Scienze Fisiche, Universit\`a degli studi di Napoli ``Federico II'' and INFN, Sezione di Napoli, I-80125 Napoli, Italy}
\address[a25]{Prisma cluster of Excellence, Institute for Physics, Johannes-Gutenberg-Universit\"at Mainz, D-55099 Mainz, Germany}
\address[a26]{Rudolf Peierls Centre for Theoretical Physics, Oxford OX13PN, UK}
\thanks{Corresponding author: gudrun@mpp.mpg.de}

\begin{abstract}
We present an update of the Binoth Les Houches Accord~(BLHA) to standardise
the interface between Monte Carlo programs and codes providing one-loop
matrix elements.

\end{abstract}



\end{frontmatter}


\section{Introduction}
\label{intro}
The past years have seen an enormous progress in the development of programs
providing next-to-leading order (NLO) corrections for multi-particle final states.
This is due to new developments concerning the calculation of one-loop
amplitudes as well as important progress on the Monte Carlo side to account
for real radiation at NLO.  The modular structure of NLO calculations allows
to share the tasks between a ``One-Loop Provider~(OLP)", providing the
virtual corrections, and a Monte Carlo program~(MC) taking care of all the
parts which do not involve loops.  To facilitate the cross-talk between those
two engines, a standard interface has been worked out during the workshop on
Physics at TeV Colliders at Les Houches in June 2009, called the ``Binoth Les
Houches Accord~(BLHA)"~\cite{Binoth:2010xt}.

Meanwhile, the use of this interface~\cite{Gleisberg:2007md, Gleisberg:2008ta,
   Berger:2010zx,Bern:2011ep, Frederix:2010ne,Hirschi:2011pa,Cullen:2011ac,
  Badger:2012pg,Badger:2012pf,Alioli:2011as,Re:2012zi, 
  Cullen:2012eh, Luisoni:2013cuh, Bern:2013gka, Hoeche:2013mua} and
further developments in 
OLP~\cite{Berger:2010zx,Bevilacqua:2011xh,Hirschi:2011pa,Cullen:2011ac,Cascioli:2011va,Badger:2012pg,Actis:2012qn} 
and
MC~\cite{Gleisberg:2008ta,Bahr:2008pv,Alioli:2010xd,Alwall:2011uj,Campbell:2012am,Alioli:2012fc,Ritzmann:2012ca,Frederix:2008hu,Frederix:2009yq} 
codes have brought up the necessity to
extend it with further options.  The aim of this article is to provide a
public document where an update of the BLHA is proposed and conventions are
defined to pass parameters, calculational schemes etc., and to return less
inclusive information, such as matrix elements which are not summed over all
colours and helicities.

\section{Existing features of the interface}
\label{sec:blha1}

We do not aim at an exhaustive description of the complete framework of the original interface
here, referring to~\cite{Binoth:2010xt} for more details.
However, we sketch the main features any extension will build upon.

The  interaction between an OLP and a
MC proceeds in two phases: the {\it pre-runtime phase} 
where order and contract files are established, and the actual {\it runtime}
phase.
In the {\it pre-runtime phase}, the MC creates a file
called {\tt order file} containing information about the
setup and the 
subprocesses it will need from the OLP to perform the computation. 
A subprocess can be
either a partonic subprocess or a component thereof (e.g.~a specific helicity 
amplitude or a colour partial amplitude).
The particles are identified by specifying their particle data group (PDG) code. 

\begin{figure}[htb]
\begin{center}
\includegraphics[width=13.cm]{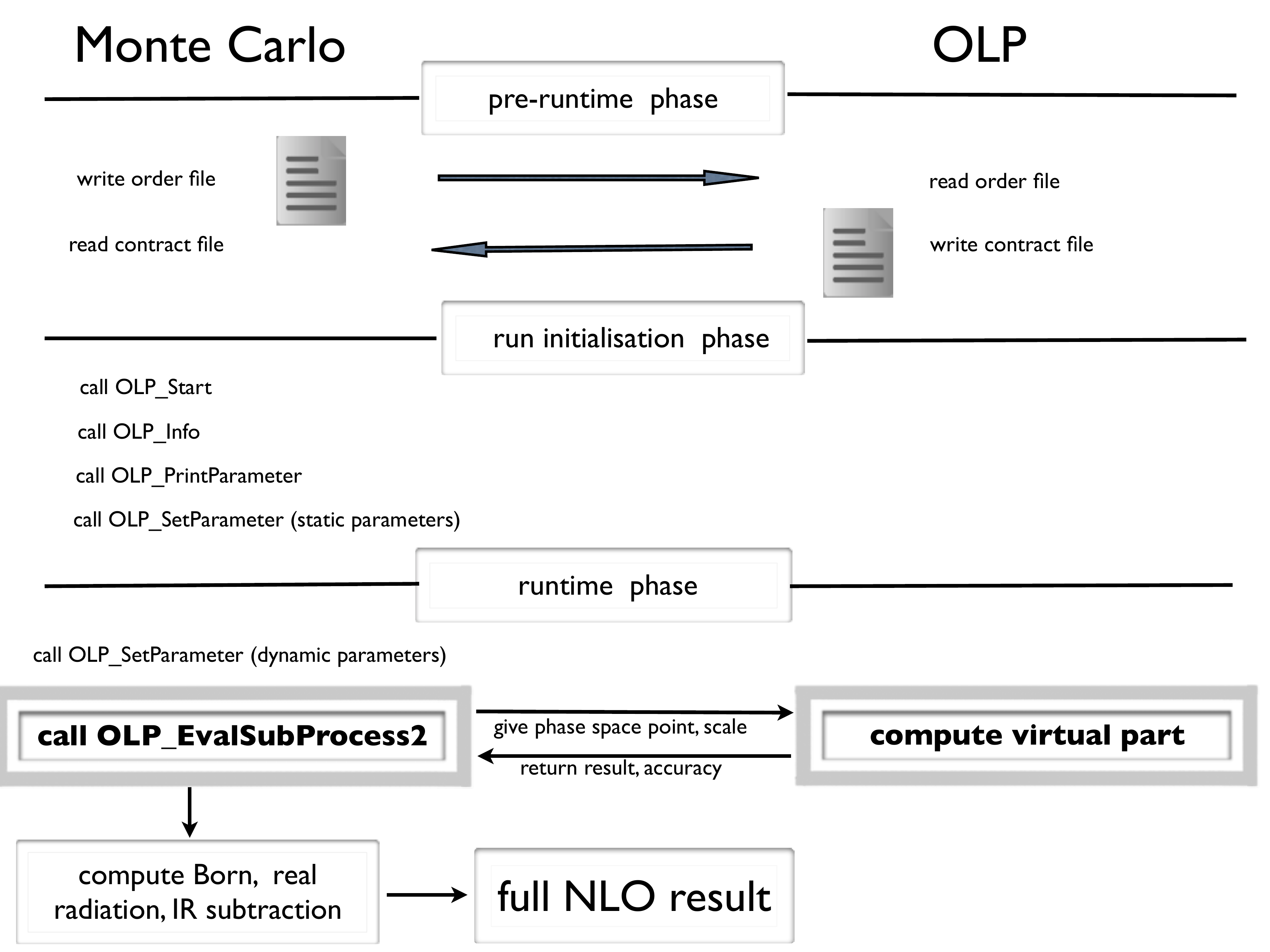}
\caption{Illustration of an interplay between Monte Carlo program and One Loop Provider~(OLP).
  In the pre-runtime phase of the interface, the OLP receives an {\tt order
  file} placed by the MC and checks availability of the contents. 
  Then it returns a {\tt contract file} to the  MC where the contents are confirmed 
  if available. 
  At runtime, the Monte Carlo program provides Born,
  real radiation part minus infrared subtraction terms and integrated
  subtraction terms. The OLP provides the virtual amplitude for each phase
  space point.  The phase space integration is done by the Monte Carlo
  program.}
\label{Fig:MC-OLP}
\end{center}
\end{figure}

A flowchart of the setup between 
Monte Carlo program  and One Loop Provider 
(where the new functions defined in BLHA version~2 
are included already) is shown in Fig.~\ref{Fig:MC-OLP}.
As an example, an order file  written 
by {\sc Sherpa} for $pp\to (Z\to e^+e^-)+1$\,jet, 
using the original standards,   is given in Fig.~\ref{fig:orderzj}.
\begin{figure}[htb]
\begin{center}
{\small 
\begin{boxedverbatim}
# OLE_order.lh
# Created by Sherpa-1.4.1

MatrixElementSquareType CHsummed
CorrectionType          QCD
IRregularisation        DRED
AlphasPower             1
AlphaPower              2
OperationMode           CouplingsStrippedOff

Z_mass                  91.188
Z_width                 2.49
W_mass                  80.419
W_width                 2.0476
sin_th_2                0.221051079833

# process list
1 -1 -> 11 -11 21
21 1 -> 11 -11 1
21 -1 -> 11 -11 -1
2 -2 -> 11 -11 21
21 2 -> 11 -11 2
21 -2 -> 11 -11 -2
\end{boxedverbatim}
}
\end{center}
\caption{Example of a {\tt BLHA1}  order file for the process $Z+$jet, created by {\sc Sherpa}.\label{fig:orderzj}}
\end{figure}

The OLP reads the order file and checks availability for each item. 
Then it returns a {\tt contract file} telling the MC what it can provide, 
and labelling the individual subprocesses.
An example contract file generated by {\sc GoSam} as a response to 
{\sc Sherpa}'s order file,
looks like the one in  Fig.~\ref{fig:contractzj}.
From this file, one can already see where an upgrade of the interface is clearly needed:
as the original interface version did not contain a 
standard way  to pass parameters,
the definition of masses and widths is marked as ``{\tt Ignored by  OLP}" in the 
contract file. Certainly the parameters are passed in the actual calculation, 
but in a non-standardised way, as an individual agreement between the particular 
MC and OLP. How to define a standard for the passing of parameters is described in 
Section~\ref{sec:setparameter}.

\begin{figure}[htb]
\begin{center}
{\small 
\begin{boxedverbatim}
# vim: syntax=olp
#@OLP GOSAM 1.0
#@IgnoreUnknown True
#@IgnoreCase False
IRregularisation DRED | OK
AlphaPower 2 | OK
sin_th_2 0.221051079833 | OK # Ignored by OLP
Z_width 2.49 | OK # Ignored by OLP
Z_mass 91.188 | OK # Ignored by OLP
W_mass 80.419 | OK # Ignored by OLP
CorrectionType QCD | OK
AplhasPower  1 | OK
W_width 2.0476 | OK # Ignored by OLP
OperationMode CouplingsStrippedOff | OK
MatrixElementSquareType CHsummed | OK
1 -1 -> 11 -11 21 | 1 3
21 1 -> 11 -11 1 | 1 4
21 -1 -> 11 -11 -1 | 1 5
2 -2 -> 11 -11 21 | 1 0
21 2 -> 11 -11 2 | 1 1
21 -2 -> 11 -11 -2 | 1 2
\end{boxedverbatim}
}
\end{center}
\caption{Example of a {\tt BLHA1} contract file for the process $Z+$jet, created by {\sc GoSam}.
As the original interface did not define a standard way how to pass parameters,
the definition of masses and widths is marked as ``{\tt Ignored by  OLP}" in the 
contract file, while the parameters are passed in a non-standardised way.
\label{fig:contractzj}}
\end{figure}

The first integer label  after each subprocess specifies that 
this subprocess contains only one component 
(if it was composed e.g.~of several helicity configurations to be evaluated separately, 
this first label would be an integer larger than one).
The second integer acts as a label for each subprocess, 
used at runtime to call the individual subprocesses.


After the contract has been ``signed", 
the communication between MC and OLP proceeds via function calls.
Signing a contract means that the OLP basically copies the order file and appends an 
{\tt OK} after each setting, separated by a ``$|$" character. 
For the requested subprocesses, the OLP answer should be the integer labels 
described above, denoting the components and the labels of the individual subprocesses.
If a setting is not supported or a subprocess is not available, the OLP should indicate this 
with {\tt Error} instead of {\tt OK} after the setting. Preferably, the message ``{\tt Error}" 
should be supplemented by a specification of the error, like {\tt unsupported flag}, 
{\tt process not available}, etc. More examples can be found in~\cite{Binoth:2010xt}.
In any case, the program should not proceed if {\tt Error} appears in the contract file.

If the contract file does not contain any {\tt Error} statements, the communication 
via function calls can be started.
In the original standard, there were only two functions which allowed the transfer of information 
between the two programs. 
One was the function {\tt OLP\_Start(char* fname, int* ierr)} which should 
be called by the MC when initializing the runtime phase. 
The character string in the first argument contains the name of the contract file. 
The integer in the second argument is set to 1 
by the function call if the contract file is accepted. 
In case of failure, the second argument is different from one, and an error message 
of the type {\tt Error: can not handle contract file} should be issued.

The second function which was already in place, but will have 
a different argument list with the new standards, is the function 
{\tt OLP\_EvalSubProcess}. 
The parameters to be passed to the {\tt OLP\_EvalSubProcess} function 
according to the original version of the interface were (in
this order):
\begin{itemize}
    \item The integer label of the subprocess (as given in the contract file).
    \item An array containing the components of the momenta. The momenta are
      placed in a one dimensional array, where physical
      scattering kinematics is used, i.e.~$k_1+k_2=k_3+\dots +k_m$.  
      For each particle, the
      kinematics is specified by a 5-tuple: $(E_j,k_j^x,k_j^y,k_j^z,M_j)$.  
      Thus a full $m$-particle event is specified by an array of 
      $5 \times m$ double precision numbers
      filled with the $m$ 5-tuples, ordered by the particle labels.
    \item The renormalisation scale, $\mur$, as a double precision number, or
      an array of scales, if different scales need to be passed.
    \item The strong coupling $\as(\mur)$, where 
    $\as(\mur)=1$ can be used to
      indicate that the {\tt MC} multiplies the returned values with the
      adequate coupling constants.  
    \item The array where the computed results are returned.
\end{itemize}
The returned array is expected to contain at least
four real-valued double precision numbers
\begin{center}
\begin{boxedverbatim}
PoleCoeff2, PoleCoeff1, PoleCoeff0, BornSquare
\end{boxedverbatim}
\end{center}
which correspond to the colour- and helicity-summed (resp. averaged 
for the initial state) terms $A_2$, $A_1$, $A_0$, $|\textrm{Born}|^2$.

The conventions for the overall prefactor are as in the original proposal~\cite{Binoth:2010xt}
and are briefly repeated here.
The general structure of the virtual correction is given by
\begin{equation}
\label{LaurentSeries}
\mathcal{I}\!\(\{k_j\},{\rm R.S.},\mur^2,\as(\mur^2),\alpha,\dots\) =
C(\epsilon) \left(\frac{A_2}{\epsilon^2} + \frac{A_1}{\epsilon}+ A_0 \right),
\end{equation}
where R.S. defines the infrared regularisation scheme.
The Laurent coefficients $A_j$ are real-valued. 
The overall constant is given by
\begin{eqnarray}
C(\epsilon) = \frac{(4\pi)^\epsilon}{\Gamma(1-\epsilon)}
\(\frac{\mu^2}{\mur^2}\)^\epsilon = (4\pi)^\epsilon
\frac{\Gamma(1+\epsilon)\Gamma(1-\epsilon)^2}{\Gamma(1-2 \epsilon)}
\(\frac{\mu^2}{\mur^2}\)^\epsilon .
\end{eqnarray}


\section{New features of the interface}
\label{sec:interface}

The new features should serve to pass more  detailed information
between the MC and the OLP. However, it should always be kept in mind that the 
general setup is such that the MC is steering the calculation. The MC orders the virtual amplitude 
for the given settings, and gets back the result and a relative accuracy informing about the quality of
the result. The interface does not foresee a setup where the OLP can cause changes in the 
MC settings.

\medskip

For the {\it pre-runtime phase}, 
we define a number of new keywords 
to allow for more options in the order/contract files.
The valid keywords are listed in Appendix~\ref{sec:newkeywords}.

As the new standards are not  backwards compatible, 
we propose to place the keyword	{\tt InterfaceVersion}, which can take the values
{\tt BLHA1} or {\tt BLHA2}, on top of the order file. 
This way, if the OLP does not support one or the other, it can issue an error message and stop 
without proceeding further.

Once order and contract files are established, one can distinguish two phases: 
The {\it run initialization phase} and the actual {\it runtime phase}. 
The functions which are used for the communication between MC and OLP 
before exchanging phase space points and results are the following (described in more detail in the 
subsections below and listed in Appendix~\ref{sec:newfunctions}):
\begin{itemize}
\item {\tt OLP\_Start(char* fname, int* ierr)}: same as in {\tt BLHA1}.
\item {\tt \small OLP\_Info(char olp\_name[15],char olp\_version[15],char message[255])}:\\
the function serves to keep track of the type and version of the OLP which has been used,
and to encourage proper citation. 
The arguments are the name of the OLP, the version, and a string which  
contains information about
the relevant publications, for example the bibtex identifier.
\item {\tt OLP\_SetParameter(char* para, double* re, double* im, int* ierr)}:
This function is used to define static parameters at the beginning of a run, 
and to exchange dynamic parameters at runtime.
\item {\tt OLP\_PrintParameter(char* filename)}: 
prints out a list of the actual parameter settings to the file {\it filename}.
\end{itemize}

At {\it runtime}, the OLP returns the values for the virtual amplitude to the 
MC via the function  {\tt OLP\_EvalSubProcess2}.
As compared to the original function {\tt OLP\_EvalSubProcess} described in Section \ref{sec:blha1},
the new function {\tt OLP\_EvalSub-}\-{\tt Process2} 
does not contain the passing of coupling constants 
anymore, as their values are now passed separately, using {\tt OLP\_SetParameter}. 
It also has a new argument  appended 
which is useful to assess the accuracy of the result returned by the OLP.
With the new argument list, the function is not backwards compatible 
with the original standard. 
Therefore, to avoid confusion with different versions,
the new function is called  {\tt OLP\_EvalSubProcess2}.

{\small 
\fbox{{\tt OLP\_EvalSubProcess2(int* i,double* pp,double* mu,double* rval,double* acc)}}
}

The arguments are:
\begin{itemize}
\item  i: pointer to a (one element) array with the label of the subprocess as given in the contract file
\item  pp: pointer to an array of momenta, conventions $(E_j,k_j^x,k_j^y,k_j^z,M_j)$
\item  mu: pointer to the renormalisation scale 
\item rval: pointer to an array of return values
\item acc: pointer to a one element array with the outcome of the 
OLP internal accuracy check (see Section \ref{sec:unstable}).
\end{itemize}
Note: originally, the argument list of {\tt OLP\_EvalSubProcess}  contained both
$\mur$ and $\alpha_s(\mur)$. 
However, $\alpha_s(\mur)$ can now be set using
the new function 
{\tt OLP\_SetParameter}  to pass also dynamical parameters. 
This setup is also clearer for mixed (e.g.~QCD-EW) 
corrections or corrections where $\alpha_s$ at different scales should be used 
within the same calculation.

The default length of the array {\tt rval} is four, containing the Laurent coefficients 
$A_2,A_1,A_0$ and $|\rm{Born}|^2$. 
For the case where colour/spin correlated 
matrix elements are returned, the array {\tt rval} must be longer. 
Details about the labelling conventions for such cases are given in 
Section~\ref{sec:colhel}.

The last argument {\tt acc} should return information about the OLP internal accuracy check(s), 
denoting the relative accuracy 
of the virtual amplitude at this phase space point as estimated by the OLP.
OLPs which only provide a binary stability test will return 0. for 
{\it passed}, or a large number, say $10^5$, for {\it failed}.
More details about unstable phase space points are given in Section \ref{sec:unstable}.

The list of defined functions is given in Appendix~\ref{sec:newfunctions}, 
where the {\tt C/C$^{++}$} version (with pointers in the argument list) is given.
The {\tt C$^{++}$} version with function calls by reference as well as 
a {\tt Fortran (2003)} module for binding with {\tt C/C$^{++}$} can be found at
{\tt http://phystev.in2p3.fr/wiki/2013:groups:sm:blha:api}.

\subsection{Passing parameters}
\label{sec:setparameter}

In the first version of the interface, the standard only 
allowed  to pass a fixed amount of  
information at the level of the order/contract files.
However, to be able to pass also dynamical parameters like running masses,
and to have more flexibility in the  definition of individual parameters, 
we suggest the following extension.

Parameters can be passed by the function

\fbox{{\tt OLP\_SetParameter(char* para, double* re, double* im, int* ierr)}}
 
where the first argument is a (pointer to a) string serving as a keyword 
for the parameter to be set, followed by two double precision numbers
so that complex parameters can also be passed (in case of real parameters, 
the second double is zero). The integer in the fourth argument 
is set by the OLP to tell the MC whether the setting of the parameter 
was successful.

{\tt ierr=1} means the parameter has been set successfully, \\
{\tt ierr=0} means failure: issue an error message, \\
{\tt ierr=2} means that the parameter is unknown 
or the setting is ignored (for example because it is irrelevant 
for the considered case), but the MC program should proceed.

The function {\tt OLP\_SetParameter} can be called at runtime, 
for every phase space point, 
if used to define a dynamic parameter. Obviously it can also be called once 
(for each particular parameter that needs to be passed) 
if this is a static parameter needed only at the run initialization phase.

Further, we propose a routine 
{\tt OLP\_PrintParameter(char* filename)} giving out a list of 
the actual  parameter settings used in the calculation,
where {\tt filename} is the name of the output file. 
The intention of this  function is just to inform the user, 
consistency between the parameters will not be checked.
The output format of {\tt OLP\_PrintParameter} should be\\
{\tt ParameterName   \quad    Value  \quad     State},\\
where the separator is a space, and 
{\tt Value}\, can be complex, denoted by {\tt (real part, imaginary part)} in case it is complex.
{\tt State} can serve  to distinguish the parameters set by {\tt OLP\_SetParameter} or
defined by {\tt Model} from fixed internal ones and 
is optional (as this info may not be readily available in all programs).

\subsection{Defining the model}
\label{sec:setmodel}

We distinguish two alternative ways of model definition, which we will denote by 
``keyword model" respectively ``{\tt UFO} model" in the following.

Model definitions offer the possibility to define some global settings 
in the order file, which are intrinsic to the model (e.g. SM, MSSM), which 
is used.
This is done using the required keyword {\tt Model}.
For example, {\tt Model: SMdiag}  should set the CKM matrix to unity globally. 

In the ``keyword model" setup, 
the parameters that need to be set within a certain model 
are passed via PDG codes~\cite{Beringer:1900zz} and keywords 
with naming
conventions as specified in Fig.~\ref{tab:keywords:static} for the Standard
Model. The numbers in parenthesis after {\tt mass} and {\tt width}  denote
the particle's PDG code.

\begin{figure}[htb]
\begin{tabular}{|l|l|}
\hline
keyword & parameter\\
\hline
{\tt mass(5)} & b quark mass \\
{\tt mass(6)} & top quark mass \\
{\tt width(6)} & top quark width\\
{\tt sw2}& $\sin^2\theta_w$\\
{\tt vev}& SM vacuum expectation value\\
{\tt Gf} & $G_{\rm{Fermi}}$\\
{\tt VV12}& $V_{ud}$\\
$\vdots$ & \\
\hline
\end{tabular}
\caption{List of keywords to define parameters to be passed by the function {\tt
OLP\_SetParameter}.}
\label{tab:keywords:static}
\end{figure}

In the ``{\tt UFO} model" setup, the parameters are defined in {\tt UFO} (Universal Feynrules
Output)~\cite{Degrande:2011ua} format, which is particularly useful for
calculations beyond the Standard Model.
The import of the {\tt UFO} model file should be specified in the {\tt order file} 
by \\{\tt Model ufo:/path\_to\_ufo\_model-directory/}.

The {\tt UFO}  format also provides human readable name attributes for the 
model parameters, as well as the 
SLHA identifiers~\cite{Skands:2003cj, Hahn:2006nq, Allanach:2008qq}. 
OLPs which support the import of {\tt UFO} model files typically also support 
the name attributes. 
The {\tt UFO} model setup entails the use of a SLHA parameter card to initialize the runtime phase.
This requires an additional keyword {\tt ParameterCard}, followed by the path to the SLHA parameter card,
to be placed into the order file when using the {\tt UFO} model setup. 
The parameters which are set by reading in the SLHA parameter card do not need to be set 
again by {\tt OLP\_SetParameter}. However, {\tt OLP\_SetParameter} needs to be used at 
runtime for the dynamic parameters. 
In this case the SLHA block name should appear as a prefix prepended to the parameter name, 
in the form  
{\tt <BlockName>\&\&<ParamName>}.
To avoid confusion, this requires that the characters `{\tt \&\&}' should never appear in 
any block or parameter name.

Note that we do not require the capability to import {\tt UFO} model files to belong to the 
``minimal standard". However, at least one of the two ways described above 
to define the model parameters should be implemented to 
comply with the standard.



\subsection{Treatment of unstable phase-space points}
\label{sec:unstable}

In a complex multi-leg one-loop calculation, some of the phase space points 
may lead to large numerical cancellations within the virtual amplitude, 
leading to  a result which is not trustworthy at this particular point.
It is the task of the OLP to make precision tests, such that it 
can trigger some rescue procedure in case the phase space point is 
found to be problematic.
In order to steer precision requirements from the order file, 
we introduce the following:
\begin{itemize}
\item The OLP should pass information about
the quality of the virtual result at a given phase space point back to the MC, 
by means of the function {\tt OLP\_EvalSubProcess2}.
The last argument of {\tt OLP\_EvalSubProcess2} should be a (double precision) number 
``{\tt acc}", 
giving the relative accuracy the OLP attributes to the  
finite part of the loop amplitude at this phase space point. 
Note that this value will depend on the type of internal stability test(s) made by the OLP. 
The assessment of the internal ``relative accuracy"
can therefore vary  between different OLPs and even within the same OLP if it can switch between 
different types of stability tests.
However, it allows the MC to give a more reliable overall accuracy estimate 
for the cross section 
than if it had just a ``pass" or ``fail" information from the OLP side.
\item The user should be able to specify the overall accuracy he would like to reach 
(this is typically done in the MC runcard).
For this purpose there should be an (optional) keyword in the order file 
which specifies the target accuracy, called {\tt AccuracyTarget}, denoting a relative accuracy. 
The OLP can use it to decide whether an internal rescue system should be triggered if 
its stability test outcome gives a value larger than {\tt AccuracyTarget}.
If {\tt AccuracyTarget} is not specified, the OLP will use its internal settings 
to decide if a rescue system should be triggered.

It could be that a particular OLP is not able to 
provide a number for the relative accuracy, but only 
can deliver a ``binary" test outcome {\it passed} or {\it failed}. 
In this case, {\tt acc=0.0} stands for {\it passed}, and in the case of {\it failed},
{\tt acc} can be set to a large value, for example $10^5$.
In order to allow the MC to distinguish the ``binary test" return value for {\it passed}
from a ``quantitative test" return value for {\it passed}, we  can 
require that quantitative stability tests will return a perfect accuracy as 
working precision (typically 1d-17, 1d-34), while binary stability tests return exactly 0.0.
\item There should be the possibility that the OLP prints points classified as 
``unstable" to a file for monitoring purposes.
Therefore we propose an 
 optional keyword {\tt DebugUnstable} in the order file which 
can be used to keep information about points classified as ``unstable".
In the case of 
{\tt DebugUnstable  True}, the phase space point should be printed to a file to allow 
further diagnostics.
The format of such a debug file can be defined internally within the OLP.
The threshold which defines whether a point is classified as ``unstable" 
(after eventual rescue attempts) 
does not need to be equal to the value of {\tt AccuracyTarget}, but this should be the default.
In any case, {\tt DebugUnstable} is mostly for OLP developers for monitoring purposes.
\end{itemize}
We emphasize again that after all, it is left to the MC to decide what to do with the phase space point, 
based on {\tt acc} returned by {\tt OLP\_EvalSubProcess2}.

\subsection{Different powers of coupling constants, merging different jet
  multiplicities} 
\label{sec:multiplicities}
So far, the interface was tailored to NLO calculations for a fixed jet
multiplicity, and focused on QCD corrections rather than electroweak
corrections.
However, mixed QCD-EW corrections, or expansions in parameters other than
$\alpha_s$ or $\alpha$, require a more flexible scheme to define the desired
orders in coupling constants.

In addition, recent developments~\cite{Hoeche:2012yf, Gehrmann:2012yg,
  Frederix:2012ps, Hamilton:2012rf, Platzer:2012bs,Lonnblad:2012ix, Luisoni:2013cuh} 
  propose a merging method
for matched NLO predictions with varying jet multiplicity.  In order to
calculate merged samples, the Monte Carlo program needs to ask the OLP for
one-loop matrix elements with different jet multiplicities and therefore
different powers of the coupling constant.


\begin{figure}[htb]
\begin{center}
{\small 
\begin{boxedverbatim}
CouplingPower QCD  2
# process list 2j
1 1 -> 1 1
1 -1 -> 1 -1
1 -1 -> 2 -2
1 -1 -> 21 21
21 21 -> 21 21

CouplingPower QCD  3
# process list 3j
1 1 -> 21 1 1
1 -1 -> 21 1 -1
1 -1 -> 21 2 -2
1 -1 -> 21 21 21
21 21 -> 21 21 21

CouplingPower QCD  4
AmplitudeType   Tree
# process list 4j
21 21 -> 21 21 21 21
1 -1 -> 21 21 21 21
1 -1 -> -2 2 -2 2
1 -1 -> -1 1 -1 1
21 21 -> 21 21 21 21
\end{boxedverbatim}
}
\caption{Example of the part of an order file containing different settings for different sets of
subprocesses.}
\label{fig:setsubprocess}
\end{center}
\end{figure}
These situations can be accounted for by allowing different settings for
different subprocesses, see Fig.~\ref{fig:setsubprocess}.
A setting is valid
until it is explicitly overwritten.  This setup can be used for merged
samples as well as mixed QCD/EW corrections.  It can also be used to pass
additional information referring only to particular subprocesses, as
indicated e.g.~by {\tt AmplitudeType Tree}.  
In the latter case, the return value {\tt rval} is a one element array 
containing the Born matrix element squared, $|\rm{Born}|^2$.


\subsection{Extras}
\label{sec:extra}

The keyword {\tt Extra} can be used to write special requirements 
relevant to the OLP into the order file.
If there is a way to  generate the order file for the OLP from the MC
run card, the MC will write them to the order file, but otherwise 
ignore them.
The difference between {\it optional keywords} and keywords preceded by 
the ``{\tt Extra}" flag is that optional keywords have defined names
(as listed in Appendix \ref{sec:optional}),  while keywords 
flagged by ``{\tt Extra}" can be very OLP specific and do not have a 
standard name.

For example, requirements concerning a colour expansion,
or passing only particular helicity configurations, could be put under the 
{\tt Extra} flag.
The {\tt Extra} flag can also be used to define that MC and OLP should use 
the {\sc MiNLO}~\cite{Hamilton:2012np} procedure, if the latter is available on both sides.

With regards to overall averaging and symmetry factors, 
the default is that amplitudes are summed over final state colours and polarisations
and averaged over initial state colours and polarisations, and that symmetry factors for 
identical final state particles are included.
A possibility to change this could look as follows: 

\begin{center}
\begin{boxedverbatim}
Extra HelAvgInitial False
Extra ColAvgInitial False
Extra MCSymmetrizeFinal False
\end{boxedverbatim}
\end{center}


\subsection{Electroweak corrections}

In the case of electroweak (EW) corrections, it is of particular importance 
to check the consistency of the parameters, for example the relation 
between $M_Z,M_W$ and $\sin^2\theta_w$. 
The scheme is set in the order file by the keyword {\tt EWScheme}, 
which can take the values 
{\tt alphaGF} (also known as $G_{\mu}$-scheme),
{\tt alpha0}, {\tt alphaMZ},  {\tt alphaRUN}, {\tt alphaMSbar}, {\tt OLPDefined}.
Then the parameters are set  using {\tt OLP\_SetParameter}.
The OLP imports these parameters. The integer in the last argument 
only indicates if the import was successful or not.
It is left to the  user to ensure consistency of the parameters within the given scheme.



\subsection{An example order file}

An example for an order file is shown in Fig.\,\ref{fig:orderttjets}.
The lines following the keyword {\tt Extra} stand for any extra information which 
may be relevant to the OLP (which, in this example, is parsed from  the {\sc Sherpa} runcard 
to the order file).

\begin{figure}
\begin{center}
{\small
\begin{boxedverbatim}
# OLE_order.lh
# Created by Sherpa-2.0.0

InterfaceVersion         BLHA2
Model                    SMdiag
AmplitudeType            CHsummed
CorrectionType           QCD
IRregularisation         DRED
WidthScheme              ComplexMass
EWScheme                 alphaGF
AccuracyTarget           0.0001
DebugUnstable            True
Extra                    Line1
Extra                    Line2

# process list
CouplingPower QCD        2
CouplingPower QED        0
1 -1 -> 6 -6
-1 1 -> 6 -6
21 21 -> 6 -6
CouplingPower QCD        3
CouplingPower QED        0
1 -1 -> 6 -6 21
1 21 -> 6 -6 1
-1 1 -> 6 -6 21
-1 21 -> 6 -6 -1
21 1 -> 6 -6 1
21 -1 -> 6 -6 -1
21 21 -> 6 -6 21
\end{boxedverbatim}
}
\end{center}
\caption{Example of an order file for $t\bar{t}+0,1$\,jets produced by {\tt Sherpa-2.0.0 }. 
The expressions {\tt Line1,Line2} following the keyword {\tt Extra} denote any extra 
information which may be relevant for the OLP.}
\label{fig:orderttjets}
\end{figure}


\subsection{Colour- and spin correlated tree amplitudes}
\label{sec:colhel}

Going beyond the generic case of colour and helicity summed matrix elements, 
it becomes difficult to satisfy special needs of different programs with one
global standard.  
Results for polarized 
amplitudes in general depend on phase conventions, 
and approximations in a colour expansion,  
like e.g. ``leading colour", will very likely be defined differently in different codes.
Nonetheless, the interface in principle allows to pass very specific information using the 
{\tt Extra} flag. 
However, the details probably remain to be implemented and tested individually  
between specific programs before trying to establish any standards.

Below we only focus on a  particular example  going beyond {\tt CHsummed}, which is
passing colour- and spin correlated tree amplitudes. 
The definitions are oriented at the case where they are used for the construction of NLO
subtraction terms in the Catani-Seymour formalism~\cite{Catani:1996vz}.
%

The MC can ask for colour- or spin correlated tree matrix elements by writing
one (or several) of the following keywords into the order file:

\begin{center}
\begin{boxedverbatim}
AmplitudeType ccTree   # colour correlated tree amplitude
AmplitudeType scTree   # spin correlated tree amplitude
\end{boxedverbatim}
\end{center}
The defaults for {\tt AmplitudeType} are {\tt Loop} and {\tt CHSummed}.
If  only {\tt Tree} or {\tt Loop} are specified, this automatically 
means {\tt CHSummed}.

The new setup to define individual settings for blocks of subprocesses can be used
to call certain subprocesses  twice: one for the colour/spin correlated tree case
and one for the loop case.
An example is given by

\begin{center}
\begin{boxedverbatim}
AmplitudeType ccTree
1 -1 -> 21 1 -1
21 21 -> 21 21 21

AmplitudeType Loop
1 -1 -> 21 1 -1
21 21 -> 21 21 21
\end{boxedverbatim}
\end{center}

If the amplitude returned by the OLP is not colour/polarisation  summed, 
the values to be returned by {\tt OLP\_EvalSubProcess2} form a matrix in 
colour respectively Lorentz space. 
To pass the values in an unambiguous way, it is  necessary to 
define the order in which the matrix elements are written
into the array {\tt rval} returned by the OLP.


\subsubsection*{Colour correlations}

The colour correlated matrix elements
\begin{equation}
C_{ij} = \langle {\cal M}| {\mathbf T}_i\cdot {\mathbf T}_j |{\cal M}\rangle
\end{equation}
can be defined to be real valued quantities and independent of the particular colour
basis chosen. Note that ${\mathbf T}_i\cdot {\mathbf T_j}$ is
symmetric under exchange of $i$ and $j$. Further, ${\mathbf T}_i^2 =
C_i {\mathbf 1}$ with $C_i=C_F$ if the leg $i$ belongs to the
${\mathbf 3}$ or $\bar{{\mathbf 3}}$ representation of ${\rm SU}(3)$,
and $C_i=C_A$ if the leg belongs to the ${\mathbf 8}$
representation, and $\langle {\cal M}| {\mathbf T}_i\cdot {\mathbf
  T}_i |{\cal M}\rangle = C_i \,\langle {\cal M}|{\cal M}\rangle =
C_i\,|{\cal M}|^2$ is just proportional to the tree level amplitude
squared. 

The minimal information that needs to be passed for 
colour correlated matrix elements 
are the upper non-diagonal $n(n-1)/2$ elements of an $n\times n$ matrix,
where $n$ is the number of legs attached to the process of
interest. Entries belonging to non-coloured legs should be ignored.

For a process which is flagged as \texttt{AmplitudeType ccTree} the
OLP should (through {\tt rval} returned by \texttt{OLP\_EvalSubProcess2})
return an array of length $n(n-1)/2$ such that the element at
position $i+j(j-1)/2$ (counting external legs and elements in the
array starting from zero) contains the result for $C_{ij}$ with $i<j$.




Here and in the following we assume that $i$ labels the emitter and $j$ the spectator.

\subsubsection*{Spin correlations}

When gluons are present at the Born level, the spin correlated Born amplitude is obtained by
Lorentz-contracting the corresponding Born terms with the gluon polarization
vectors. 
It can be shown that these amplitudes can be written in terms of 
\begin{equation}
S_{ij} = \langle {\cal M}_{i,-} |{\mathbf T}_i\cdot {\mathbf T}_j |{\cal M}_{i,+}\rangle \, ,
\end{equation}
with
\begin{eqnarray}
&&\langle {\cal M}_{i,-} |{\mathbf T}_i\cdot {\mathbf T}_j |{\cal M}_{i,+}\rangle =\\
&&\sum_{\lambda_1,...,\lambda_{i-1},\lambda_{i+1},...,\lambda_n}
{}_C\langle {\cal M}_{\lambda_1,...,\lambda_{i-1},-,\lambda_{i+1},...,\lambda_n} |
{\mathbf T}_i\cdot {\mathbf T}_j | 
{\cal M}_{\lambda_1,...,\lambda_{i-1},+,\lambda_{i+1},...,\lambda_n}\rangle_C \, .\nonumber
\end{eqnarray}
Here, $|\rangle_C$ indicates an object which is a vector in colour
space only, and we have made explicit the spin dependence by the
subscripts $\lambda$ and the fixed $\pm$ polarization of the gluon $i$
of interest. Depending on the organisation of the subtraction terms,
non-trivial colour correlations $i\ne j$ may be needed. Since
$i$ is defined to be the gluon index, we have 
$S_{ij}\ne S_{ji}$ and $S_{ij}\in {\mathbb C}$, so in general the
communication of a complex $n\times n$ matrix is required. As for the
plain colour correlated matrix elements, entries corresponding to
non-coloured legs or coloured legs for which the infrared singular limit
does not involve spin correlations ({\it i.e.}  everything but gluons
for all practical purposes) should be ignored.

For a process flagged as \texttt{AmplitudeType scTree} the OLP should
(through the standard \texttt{OLP\_EvalSubProcess2} method)  return
an array of length $2n^2$, such that the elements at positions $2 i+2 n
j$ and $2 i+2 n j + 1$ (counting external legs and elements in the
array starting from zero) contain $\text{Re}(S_{ij})$ and
$\text{Im}(S_{ij})$, respectively.

A consistent setup of spin correlations in this way further requires
the MC code to obtain the gluon polarization vector
$\epsilon_+^\mu(p,q)$ as used by the OLP, given the gluon momentum $p$
and a reference vector $q$. 
The gluon polarisation vectors 
can be passed via a dedicated function 
\begin{center}
\begin{boxedverbatim}
OLP_Polvec(double* p, double* q, double* eps)
\end{boxedverbatim}
\end{center} 
where {\tt p} and {\tt q} are arrays of size $4$, denoting the momentum $p$ of gluon $i$
and the corresponding reference vector, 
and {\tt eps} is an array of
size $8$ containing the four complex components of 
$\epsilon_\pm^\mu(p,q)=\pm \frac{1}{\sqrt{2}}\frac{\langle q^\mp|\gamma^\mu|p^\mp\rangle}{\langle
q^\mp|p^\pm\rangle}$, 
in a form where real and imaginary parts are alternating.




Returning all possible colour and spin correlations could potentially become
a significant overhead; we leave it to the OLP code to support the
calculation of a single correlator.  Information about the correlator to be
returned could be passed via the \texttt{OLP\_SetParameter} method right
before the \texttt{OLP\_EvalSubProcess2} call in question.

Obviously, if one of the contractors cannot provide the detailed colour or
helicity information requested, the calculation should exit at the stage of
{\tt OLP\_Start}.

\subsection{Restrictions such as diagram filters, exploitation of special symmetries, etc.}

The keyword {\tt ExcludedParticles} in the order file can be used to remove
unwanted particles from the code generation.  Restrictions like confining the
set of diagrams to resonant diagrams only can be set in the MC input card.
OLP specific restrictions can be imposed using the keyword {\tt Extra}.

\section{Conclusions}

This writeup summarises the update of the  standard interface 
between Monte Carlo programs and one-loop matrix element providers
which has been 
initiated at the Les Houches 2009 workshop on Physics at TeV Colliders, 
called the ``Binoth Les Houches Accord~(BLHA)". The setup meanwhile has been 
implemented by several groups and facilitates the automation of NLO calculations.

The experience gained meanwhile with the original setup fed into the
discussion about an extension of the standards, such that the interface can
be used in a wider and more flexible context.  The outcome of the discussion
between a large number of Monte Carlo and One-Loop Providers (OLPs) is summarized in
the present document, which is intended to serve as a reference for the new
standards.  This should increase the flexibility of both Monte Carlo programs
to import virtual corrections where available and of OLPs to
team up with different Monte Carlo programs.  This is an important step
forward, as different MCs and OLPs have different focus and strengths
concerning for example the multiplicity of final states, particle masses,
electroweak corrections, BSM capability, etc.  We therefore hope that BLHA
version~2 will contribute to the goal of extending the comparison of LHC data
with theoretical results consistently beyond the leading order.

\vspace{1cm}

{\bf Acknowledgements}

We would like to thank the CERN TH/LPCC Institute on {\it SM at the LHC 2012} 
for hospitality 
and for providing a stimulating environment to discuss this interface.
We would also like to thank the Les Houches 2013 workshop organizers for providing 
a platform (e.g.~Wikipages) and again a stimulating environment to 
work out this Accord. GH would like to 
dedicate this document to Thomas Binoth for his birthday on August 16, and
thank everybody who contributed 
to the creation and implementation of the Accord, both BLHA1 and BLHA2.


\appendix
\renewcommand \thesection{\Alph{section}}
\renewcommand{\theequation}{\Alph{section}.\arabic{equation}}
\setcounter{equation}{0}

\section{List of valid keywords for order/contract file}
\label{sec:newkeywords}

\subsection{Required keywords}
\begin{description}
\item[{\tt InterfaceVersion}:] can take the values {\tt BLHA1} or {\tt
  BLHA2}.  This clarifies already in the pre-runtime phase if the new
  standards are supported by both MC and OLP.
\item[{\tt Model}:] {\tt SMdiag}, {\tt SMnondiag}, {\tt MSSM},
{\tt ufo:/abs-path-to-ufo-file/}.
For BSM standards, the {\tt UFO}~\cite{Degrande:2011ua} format is most convenient.
\item[{\tt CouplingPower QCD}:] integer which specifies the $\as$ power of the Born
  cross section. Can be used for sub-processes as well, 
  where it also refers to cross section rather than amplitude level.
\item[{\tt CorrectionType}:] the type of higher order correction. Standard
  values are {\tt QCD}, {\tt QED}, corresponding to expansions in $\as$, $\alpha$.
\item[{\tt IRregularisation}:] the infrared regularisation scheme
  used. Possible choices for QCD are {\tt CDR}, {\tt DRED}.
%
\end{description}

\subsection{Optional keywords} 
\label{sec:optional}
  
\begin{description}
\item[{\tt AccuracyTarget}:] The target accuracy the user would like to achieve. 
This is a relative accuracy the OLP can use to decide if a rescue system 
should be triggered if the result of internal stability checks returns 
a value which is larger than {\tt AccuracyTarget}.
\item[{\tt DebugUnstable}:] 
can take the values {\tt True} or {\tt False}. 
Indicates that the unstable phase space points 
should be printed to a file for further diagnostics. 
In order to decide what is classified as ``unstable  phase space point", 
the threshold given by {\tt AccuracyTarget} or an internal OLP threshold can be used.
\item[{\tt CouplingPower QED}:] integer which specifies the $\alpha$ power of the
  Born cross section. The default is zero. 
  Can also be used for other couplings 
  in the form {\tt CouplingPower gX} if the coupling {\tt gX} is defined
  (for example through an {\tt UFO} module).
   Note that {\tt CouplingPower QCD} needs to be defined explicitly, 
   while {\tt CouplingPower QED} will be set to the default value (zero) if not specified.
\item[{\tt AmplitudeType:}] the default is {\tt Loop}.  It can also
  take the values {\tt ccTree} (colour correlated tree), {\tt scTree} 
  (spin correlated tree), {\tt Tree}, {\tt LoopInduced}.
  {\tt Tree} without ``{\tt cc}" or ``{\tt sc}" implies {\tt CHsummed}.
\item[{\tt Extra}:] can be used to write special requirements relevant to the
  OLP into the order file.
  If the MC generates the order file from a run card where such requirements are specified, 
  it will copy them into the order file but otherwise ignore them.
\item[{\tt ParameterCard}:] gives the path to the SLHA parameter card if the {\tt UFO} model setup is used.
\item[{\tt MassiveParticles}:] defines a list of massive particles at the
  level of the order file, for example {\tt MassiveParticles 5\ 6}.  The
  separator is a space.  This also implies that the light-quark masses are
  set to zero.
\item[{\tt LightMassiveParticles}:] useful if mass regularisation instead of
  dimensional regularisation is used (e.g.~for electroweak corrections). It
  defines the set of particles where only $\log(m)$ terms are kept, but not
  power suppressed ones.
\item[{\tt ExcludedParticles}:] can be used to exclude particles which are
  contained by default in {\tt Model}. The particles should be listed after
  the keyword, denoted by their PDG numbers, and separated by a space.
\item[{\tt MassiveParticleScheme}:] a standard choice is {\tt OnShell}.
\item[{\tt SubdivideSubprocess}:] this flag defines whether a given process
  is represented in a split form to allow for multi-channel Monte Carlo
  sampling. 
  May be useful e.g. in the case of colour expansions.
  Can take the values {\tt True} or {\tt False} (default).
\item[{\tt EWScheme}:] used schemes (discussed in the text) can be flagged by
  the keywords {\tt alpha0}, {\tt alphaMZ}, {\tt alphaGF}, {\tt alphaRUN},
  {\tt alphaMSbar}, {\tt OLPDefined} (default).
\item[{\tt WidthScheme}:] defines the treatment of unstable particles.
  Standard values are {\tt ComplexMass}, {\tt FixedWidth}, {\tt
    RunningWidth}, {\tt PoleApprox}.

\end{description}

\subsection{List of keywords contained in the original proposal which will be dismissed}


\begin{description}
\item[{\tt MatrixElementSquareType}:] replaced by {\tt AmplitudeType},
  declared {\tt CHsummed} as default.  Originally this flag was intended to
  specify colour~($C$) and helicity~($H$) treatment.  Possible values were
  defined as {\tt CHsummed}, {\tt Csummed}, {\tt Hsummed}, {\tt
    NOTsummed}. \\ The new default {\tt CHsummed} also implies an average
  over initial state colours and polarizations.  The flag {\tt Extra} can
  be used to accommodate for the keywords {\tt HelAvgInitial}, {\tt
    ColAvgInitial}, {\tt MCSymmetrizeFinal}, which can be set to {\tt False} if the
  factors included by default should be switched off.
%
\item[{\tt ModelFile}:] the model file from which parameters have to be read.
  Keyword replaced by {\tt Model}.
%
\item[{\tt OperationMode}:] the operating mode of the OLP. This optional flag
  was intended to specify OLP-defined conventions or approximations to the
  one-loop contribution. Typical operating modes are {\tt
    CouplingsStrippedOff}, {\tt LeadingColour}, {\tt HighEnergyLimit}.
The keyword {\tt CouplingsStrippedOff} turned out to be ambiguous in the
presence of EW couplings and therefore will not be used any longer.
\item[{\tt ResonanceTreatment}:] has been replaced by {\tt WidthScheme}, as,
  for example, the complex-mass scheme also affects non-resonant propagators.

\end{description}

\section{List of defined functions}
\label{sec:newfunctions}

We list here the definitions of the functions needed for a working interface
in the {\tt C/C$^{++}$} version (with pointers in the argument list).  A {\tt
  Fortran (2003)} module for binding with {\tt C/C$^{++}$} can be found
at\\ {\tt http://phystev.in2p3.fr/wiki/2013:groups:sm:blha:api}.

\begin{boxedverbatim}
#ifndef __OLP_H__
#define __OLP_H__
     
#ifdef __cplusplus
extern "C" {
#endif
     
void OLP_Start(char* fname, int* ierr);
void OLP_Info(char[15] olp_name,char[15] olp_version,char[255] message);
void OLP_SetParameter(char* para, double* re, double* im, int* ierr);
void OLP_PrintParameter(char* filename);   
void OLP_Polvec(double* p, double* n, double* eps); // (optional)
void OLP_EvalSubProcess2(int* i,double* pp,double* mu,double* rval,double* acc);
     
#ifdef __cplusplus
}
#endif // __cplusplus
#endif // __OLP_H__

\end{boxedverbatim}

\providecommand{\href}[2]{#2}\begingroup\raggedright\endgroup

\end{document}